\documentclass[osajnl,twocolumn,showpacs,superscriptaddress,10pt]{revtex4-1}

\usepackage{amsmath,amssymb,graphicx}
\usepackage{latexsym} 
\usepackage{amsmath,amsthm}

    \usepackage{ifpdf}

\usepackage{epstopdf}
\usepackage{dcolumn}
\usepackage{bm}
\usepackage{braket}

\begin{document}

  \ifpdf
    \else
    \fi

\title{Gamow Vectors Explain the Shock ''Batman'' Profile}

\author{Maria Chiara Braidotti}
    \affiliation{Institute for Complex Systems, National Research Council (ISC-CNR), Via dei Taurini 19, 00185 Rome (IT).}
		\affiliation{Department of Physical and Chemical Sciences, University of L'Aquila, Via Vetoio 10, I-67010 L'Aquila (IT).}
    \email{mariachiara.braidotti@isc.cnr.it}
\author{Silvia Gentilini}
    \affiliation{Institute for Complex Systems, National Research Council (ISC-CNR), Via dei Taurini 19, 00185 Rome (IT).}
\author{Claudio Conti}
    \affiliation{Institute for Complex Systems, National Research Council (ISC-CNR), Via dei Taurini 19, 00185 Rome (IT).}
    \affiliation{Department of Physics, University Sapienza, Piazzale Aldo Moro 5, 00185 Rome (IT).}
    \homepage{http://www.complexlight.org}

\date{\today}

\begin{abstract}
The description of shock waves beyond the shock point is a challenge in nonlinear physics. Finding solutions to the global dynamics of dispersive shock waves is not always possible due to the lack of integrability. Here we propose a new method based on the eigenstates (Gamow vectors) of a reversed harmonic oscillator in a rigged Hilbert space. These vectors allow analytical formulation for the development of undular bores of shock waves in a nonlinear nonlocal medium. Experiments by a photothermal induced nonlinearity confirm theoretical predictions: as the undulation period as a function of power and the characteristic quantized decays of Gamow vectors. Our results demonstrate that Gamow vector are a novel and effective paradigm for describing extreme nonlinear phenomena. 
\end{abstract}


\maketitle
\noindent In physics, shock waves emerge in a wide variety of fields ranging from fluidodynamic to astrophysics, to dispersive gas dynamics \cite{Hoefer2008Piston} and plasma physics \cite{Taylor1970,Romagnani2008}, to granular systems \cite{Molinari2009}, to Bose-Einstein condensation and polaritons \cite{Dominici2015}. \cite{Hoefer2006,Wan2007,Conforti2014,EL2007,El2005,Maiden2015,Barsi2007,Sun2012} The shock waves ubiquity arises from the universal properties of hyperbolic systems of partial differential equations, which are typically present in various contexts. However the exact description of the shock wave profiles is rarely available and techniques like the Whitham approach are limited to integrable or nearly integrable systems. \cite{Crosta2012,WhithamBook} Very simplified hydrodynamic models like the Hopf equation are used in most of the cases. \cite{Gurevich73,Bronski1994,Kamchatnov2002,Gentilini2013a,WhithamBook} These methods allow the description of the wave propagation since the occurrence of the shock point, but do not provide a global and comprehensive analysis of the phenomenon. Issues that are often overlooked include the field decay after the shock, and the long term evolution (far field). In addition, peculiar features like the oscillation period of the so called undular bores (fast oscillation observed in the wave profile in dispersive systems) only remain as results of numerical calculations. One specific open question concerns the appearance of the fast oscillations in the internal part of a Gaussian beam undergoing a shock in a nonlocal medium. This is in contrast with the hydrodynamical formulation, which predicts that the bores are expected to appear in the external part, i.e., at the beam edges.  During the shock formation the beam displays a characteristic double peaked M-shape (''Batman ears'') that is also a riddle in the field of normal dispersion mode-locked laser \cite{ContiGrelu14,Chong:06,Chong:08,Kieu:09}. No analytical solution is available to describe the phenomenon. The challenge is finding new ideas and paradigms for describing wave propagation beyond the breaking point and hence cover the knowledge gap in predicting this universal shock features.\\
Recently, unnormalizable wavefunctions named nonli\-near Gamow vectors (GVs) proved to be fruitful: the shock waves are described by the eigensolutions of the so called reversed harmonic oscillator (RHO). \cite{Gentilini2015,Glauber85}\\
Here we show that this approach can be the key to solve analytically shock wave propagation in the far field. Our analysis allows to describe and analyze the development of the characteristic undular bores during the shock formation, and provides a complete description of the ''Batman ears''.\\
      
\noindent We consider a light beam with amplitude $A$ ($I=|A|^2$ is the intensity), wavelength $\lambda$, and propa\-gating in a medium with refractive index $n_0$. Let $Z$ and $X$ the propa\-gation and polarization directions respectively, then the paraxial propa\-gation equation reads as
\begin{equation}
2ik\frac{\partial A}{\partial Z}+\frac{\partial^2A}{\partial X^2}+2k^2\frac{\Delta n[|A|^2](X)}{n_0}A=0,
\label{nls}
\end{equation}
with $P_{MKS}=\int IdX$ the beam power per unit length in the transverse $Y$ direction and $k=2\pi n_0/\lambda$ the wavenumber. In a nonlocal medium the refractive index variation $\Delta n$ can be written as 
\begin{equation}
\Delta n[I](X)=n_2\int G(X-X')I(X')dX'. 
\label{Dn}
\end{equation}
The function $G$ is proportional to the Green function and it is normalized such that $\int GdX=1$. We observe that Eq.~(\ref{Dn}) originates from the solution of the Fourier heat equation for thermal nonlocal nonlinearity.\cite{Gentilini2013a}\\
We write Eq.~(\ref{nls}) in terms of the normalized variables $x=X/W_0$ and $z=Z/Z_d$ with $Z_d=kW_0^2$
\begin{equation}
i\frac{\partial\psi}{\partial z}+\frac{1}{2}\frac{\partial^2\psi}{\partial x^2}-PK(x)\ast|\psi(x)|^2\psi=0,
\label{nls2}
\end{equation}
where $\psi=AW_0/\sqrt{P_{MKS}}$, with $\Braket{\psi|\psi}=1$ denoting Hilbert space scalar product. $P$ is set as $P_{MKS}/P_{REF}$, with $P_{REF}=\lambda^2/4\pi^2n_0|n_2|$. $n_2$ is the nonlinear optical coefficient and $K(x)=W_0G(xW_0)$ is the nonlocality kernel function.

\begin{figure}[h!]
	\centering
		\includegraphics[scale=0.40]{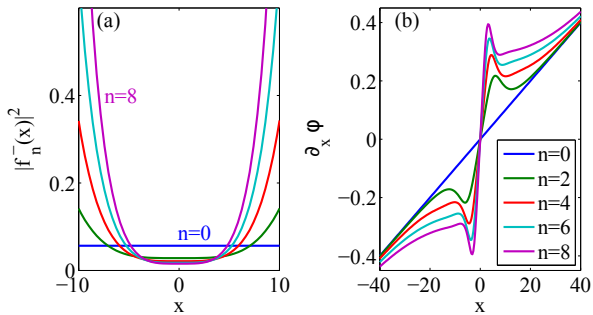}
	\caption{(color online) First five even reversed oscillator Gamow eigenstates square modulus (a) and their tilt (b), calculated as $\partial_x\varphi$, for the fundamental state and first four excited states $(n=0,\,2,\,4,\,6 \,\,\mbox{and}\,\,8)$.}
	\label{gamow_state}
\end{figure}

\noindent As shown in \cite{Gentilini2015}, in a highly nonlocal medium, i.e. when the nonlocality function width is much wider than the field intensity and the highly nonlocal approximation (HNA) is valid $[K(x)\ast|\psi|^2\simeq\kappa(x)]$, the solution of Eq.~(\ref{nls2}) is strongly linked to the eigenstates of a RHO. These states are the so called Gamow vectors, firstly introduced by Gamow in $1920$s in nuclear physics in order to describe particle decays and resonances. \cite{gamow1928quathenucdis} It can be shown that GVs can be obtained by extending the harmonic oscillator eigenfunctions in the complex plane:\cite{Chruscinski2004}
\begin{equation}
\mathfrak{f}_n^{\pm}(x)=e^{\pm i\pi/8}\left(\frac{\sqrt{\pm i \gamma}}{2^nn!\sqrt{\pi}}\right)^{1/2}e^{\mp i\frac{\gamma}{2}x^2}H_n(\sqrt{\pm i\gamma}x),
\label{autofunz_rho}
\end{equation}
where $H_n(x)$ are the $n$-order Hermite polynomials. The $\mathfrak{f}_n^{\pm}$ are discrete states belonging to a rigged Hilbert space (RHS) $\mathcal{H}^{\times}$, which is an extension of the standard Hilbert space $\mathcal{H}$. In $\mathcal{H}^{\times}$ the Khalfin theorem \cite{Khalfin57} does not hold true and exponentially decaying wavefunctions are admitted. Indeed, the eigenvalues of the RHO Hamiltonian 
\begin{equation}
\hat H_{rho}(\hat p,\hat x)=\frac{\hat p^2}{2}-\frac{1}{2}\gamma^2\hat x^2
\label{rho}
\end{equation} 
are purely imaginary numbers $E^{\pm}_n=\pm i\gamma(n+1/2)$, where $\gamma$ is the decaying coefficient of the associated classical system. For $z>0$ the eigenfunction $\mathfrak{f}^{+}_n$ is exponentially increasing while $\mathfrak{f}^{-}_n$ is decreasing. For this reason we choose the latter to describe exponential decaying dynamics when $z$ grows. Figure~\ref{gamow_state} shows the square modulus of $\mathfrak{f}^{-}_n$ and their \emph{tilt}, calculated as the $x$ derivative of the phase $\varphi$ of $\mathfrak{f}^{-}_n$, $\partial_x\varphi$, for even $n$. Notice the resemblance of these functions with the standard intensity and phase profile observed during numerical simulations and experiments in shock waves.\cite{Gentilini2013a} 

\noindent It is worthwhile to notice that we can analyze the beam evolution during wave breaking in a nonlocal nonlinear medium using GVs of RHO. When HNA holds true and writing $\psi=e^{-iPz/2\sigma}\phi$, Eq.~(\ref{nls2}) reads as (see \cite{Gentilini2015})
\begin{equation}
i\partial_z\phi=\hat H_{rho}\phi=\sum_{n=0}^{+\infty}E^-_n\Ket{\mathfrak{f}^-_n}\Bra{\mathfrak{f}^+_n}\phi.
\label{nls4}
\end{equation}
The eigenfunctions $\phi$ can be expanded in terms of Gamow eigenvectors as
\begin{equation}
\phi(x)=\sum_{n=0}^{N}\sqrt{\Gamma_n}\mathfrak{f}^-_n\Braket{\mathfrak{f}^+_n|\psi(x,0)}.
\label{gs}
\end{equation}
where $\psi(x,0)$ is the initial physical state and $\Gamma_n=\gamma(2n+1)$ are the GVs quantized decay rates. \\
RHO Gamow eigenstates have the peculiar characteristic of being the eigenvectors of Fourier transform operator. Indeed, one can observe that the reversed oscillator eigenvalue equation has the same form as its Fourier transform within a phase factor. Considering the RHO Hamiltonian in the position basis ($\hat x \rightarrow x$ and $\hat p \rightarrow -i\partial_x$) we have:
\begin{equation}
\hat H_{rho}(\hat p,i\partial_p)=\frac{\hat p^2}{2}+\frac{1}{2}\gamma^2\partial_p^2=-\hat H_{rho}(-i\partial_x,\hat x).
\end{equation}
To describe the far field with this formalism, we cannot neglect that GVs have an infinite support. Hence, to account for the spatial confinement of the experiment, we introduce the windowed Gamow vectors:
\begin{equation}
\phi^W_G(x)=\sum_{n=0}^{N}\sqrt{\Gamma_n}\mathfrak{f}^-_n\Braket{\mathfrak{f}^+_n|\psi(x,0)}\mbox{rect}_W(x),
\label{WGV}
\end{equation}  
where rect$_W(x)=0$ for $|x|>W$ and rect$_W(x)=1$ for $|x|<W$, which is the range of the spatial confinement. During evolution each Gamow component in Eq.~(\ref{gs}) exponential decay with rate $\Gamma_n$: the ground state has the lowest decay rate $\Gamma_0=\gamma$ and higher order Gamow states decay faster than the fundamental one. This allows to consider only the fundamental GV in the far field. We compute the Fourier transform $\mathcal{F}$ of the fundamental state of Eq.~(\ref{WGV}): 
\begin{equation}
\begin{aligned}
\tilde \psi(k_x)=\mathcal{F}\left(\mathfrak{f}^-_0(x)\right)&=\biggl(\frac{1}{4}+\frac{i}{4}\biggr)e^{-\frac{ikx^2}{2\gamma}}\frac{(-i\gamma \pi)^{1/4}}{W}\times  \\
&\times \left\{ -\mbox{Erf} \biggl[\frac{(\frac{1}{2}-\frac{i}{2})(k_x-W \gamma)}{\sqrt{\gamma}}\biggr] + \right. \\
& \left. + \mbox{Erf} \biggl[\frac{(\frac{1}{2}-\frac{i}{2})(k_x+W\gamma)}{\sqrt{\gamma}}\biggr] \right\}.
\label{fftdn}
\end{aligned}
\end{equation}

\begin{figure*}[t!]
	\centering
		\includegraphics[scale=0.4]{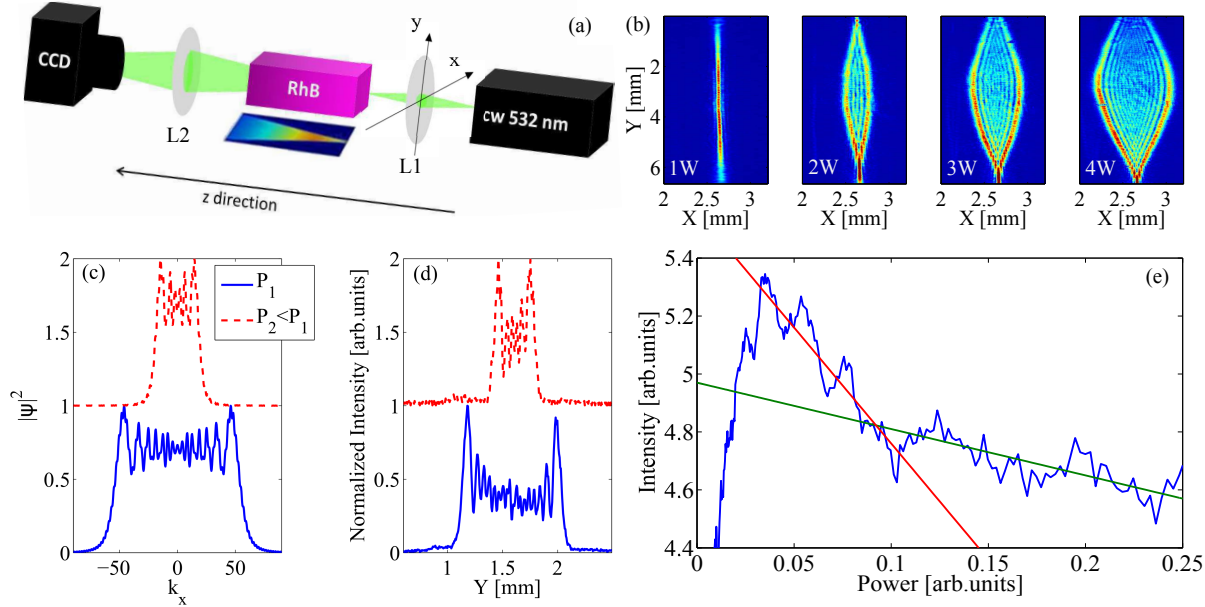}
	\caption{(color online) (a) Experimental setup scheme used to collect images of the laser beam transmitted by a RhB sample. L1 is the cylindrical lens used to make the beam elliptical and to focus it in the sample. L2 is the spherical lens used to collect the beam with a CCD camera. (b) CCD images of the light beam at different laser powers $(P_{MKS}=1\,,2\,,3 \,\,\mbox{and}\,\, 4W)$. (c) Eq.~(\ref{fftdn}) square modulus for two different power values $(P_1 > P_2)$. (d) Experimental normalized intensity profile for different power values: $P_{MKS}=2$W(dashed line) and $4$W(continuous line). (e) Intensity decays as a function of power. The coefficients of the straight lines describe the Gamow vectors decaying rates $(\gamma_1=-8.0$ and $\gamma_2=-1.6)$. Their quantized ratio is $5$ as expected from theory (see \cite{Gentilini2015}).}
	\label{experim}
\end{figure*}

\noindent Equation~(\ref{fftdn}) gives an analytical expression of the far field, which is compared below with the experiments (Fig.~\ref{experim}). Equation~(\ref{fftdn}) predicts in closed form the typi\-cal ''Batman profile'', the fact that undular bores are internal in the beam profile, and the correct scaling with respect to the power of the undulation period.\\
We validate this analysis by experiments in a nonlocal optothermal medium.  
The experimental set-up is illu\-strated in Fig.~\ref{experim}a. A continuous wave (CW) laser beam at $532$nm wavelength is focused through a cylindrical lens (L1) with focal length $f=20$cm in order to mimic a nearly one-dimensional propagation. Letting $Z$ the propa\-gation direction, the lens focuses the beam in the $X$ direction. The light is collected by a spherical lens (L2) and a Charged Coupled Device (CCD) came\-ra.
The spot dimension is $1.0$mm in the $Y$ direction and $35\mu$m in the $X$ direction. These geometrical features make the unidimensional approximation valid and allow to compare experimental results with the theoretical model.
The diffraction length in the $X$ direction is $L_{diff}=3.0$mm. A solution of Rhodamine B (RhB) and water at $0.1$mM acts as a nonlocal optical medium and is placed in a cuvette $1$mm thick in the propagation direction. RhB is a dye with a high nonlinear index of refraction $n_2$, its absorption length is $L_{abs}=1.0$mm. \cite{Gentilini2013a}\\
We collect CCD images of the beam for different powers (see Fig.~\ref{experim}b). The transverse $X$ section broadens and we observe intensity peaks (''Batman ears'') on the lateral edges of the beam, resembling the shape of Gamow vectors (see Fig.~\ref{experim}c and \ref{experim}d). The undular bores of the shock appear between the lateral peaks, in the internal part of the Gaussian beam. For low power $(P_{MKS}\leq1$W$)$ the beam profile is Gaussian. 
Figure~\ref{experim}c shows the square modulus of the far field analytically expressed in Eq.~(\ref{fftdn}) for two different power va\-lues $(P_1 > P_2)$ and we stress the remarkable agreement with the experimental results shown in Fig.~\ref{experim}d. We also observe that the experimental data exhibit a reduction in the central part of the profile. This is mostly caused by the presence of nonlinear losses (not included in the model): the thermal effect induces Rhodamine diffusion out of the highest intensity regions, which, in turn, are hence subject to a reduced absorption. \cite{Gentilini2013a}\\
Exponential decays are the major signature of Gamow states.\cite{Gentilini2015,Gentilini2015glauber,Chruscinski2004} An important aspect of our analysis is that the elliptical beam has an intensity that varies Gaussianly along $Y$. This implies that, observing a CCD image, intensity profiles at different $Y$ correspond to different powers in the one-dimensional approximation; the link between the $Y$ position and the power follow the Gaussian profile $(P_{MKS}\propto\exp{-Y^2})$. Correspondingly the expected exponential trend with respect to power can be extracted from a single picture by looking at different $Y$ positions. 
This analysis is carried out by considering a region in Fig.~\ref{experim}b at $P_{MKS}=4$W; the resulting profile versus power is shown in Fig.~\ref{experim}e: two exponential trends are clearly evident and the two straight lines corresponding to different decay coefficients are drawn (the conversion from $Y$ to $P_{MKS}$ correspond to a logarithmic scale in which exponentials are replaced by straight lines). The extracted ratio of the two decay coefficients is $5$ and hence in agreement with the expected quantized theoretical value \cite{Gentilini2015}.

\begin{figure}[h!]
	\centering
		\includegraphics[scale=0.40]{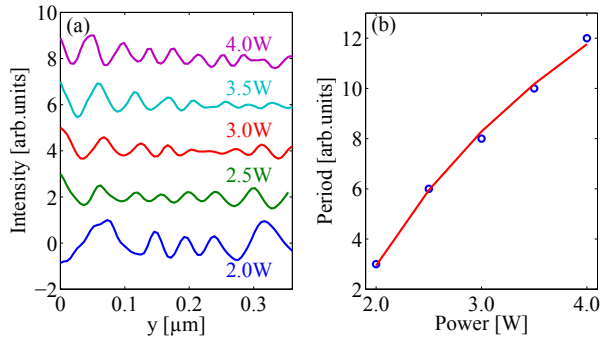}
	\caption{(color online) (a) Characteristic intensity oscillations for different power values. (b) Measured undulation period $T$ as a function of power. Continuous line is the fit function $T\propto\sqrt[4]{P}$ as expected by the theory.}
	\label{far_field}
\end{figure}

\noindent We analyze the undular bores of shock waves (see Fig.~\ref{experim}b). 
Equation~(\ref{fftdn}) predicts that the field intensity undulation period $T$ grows like $T\propto\sqrt[4]{P}$. Figure~\ref{far_field}a shows the oscillatory behavior of the far field as visua\-lized on the CCD camera when removing the collecting lens L2. Through spectral analysis we obtain the period as a function of optical power which matches the theoretical expectation as shown in Fig.~\ref{far_field}b. 

\noindent The control of extreme nonlinear phenomena is at the basis of the future developments of nonlinear physics, but requires novel theoretical tools and paradigms. 
In this article we propose a novel approach to describe the occurrence of undular bores and the M-shape (''Batman'') intensity profile during a highly nonlinear evolution ruled by the nonlocal nonlinear Shr\"{o}dinger equation. The strong nonlinearity produces shock waves, and we provide a global description of the wave breaking by new techniques from irreversible quantum mechanics. Our experiments quantitatively confirm the new theoretical scenario. We believe that this approach is not only limi\-ted to the spatial case considered here, but has an impact in temporal pulse dynamics (as for example modelocked lasers in the normal dispersion regime) and also, more in general, in the vast number of fields dealing with shock waves. Understanding the way GVs occurs during extreme nonlinear phenomena may lead to the control of these processes in fields like nonlinear optics, polaritons and ultracold physics, and to the development of novel devices including supercontinuum and X-ray generation. \\

\noindent We acknowledge support the ERC project VANGUARD (grant number 664782), the Templeton Foundation (grant number 58277) and the ERC project COMPLEXLIGHT (grant number  201766).

%

\end{document}